\documentclass[12pt]{article}
\usepackage{amssymb}
\usepackage{graphicx}

\def\be{\begin{equation}}
\def\ee{\end{equation}}

\begin{document}
\begin{center}
{\Large\bf Critical behavior of the compact $3d$ $U(1)$ theory
 in the limit of zero spatial coupling}\\ 
\end{center}
\vskip 0.3cm
\centerline{O. Borisenko$^*$}
\vskip 0.3cm
\centerline{\sl Bogolyubov Institute for Theoretical Physics,}
\centerline{\sl National Academy of Sciences of Ukraine,}
\centerline{\sl 03680 Kiev, Ukraine}
\vskip 0.3cm
\centerline{M. Gravina$^{**}$ and A. Papa$^{**}$}
\vskip 0.3cm
\centerline{\sl Dipartimento di Fisica, Universit\`a della Calabria,}
\centerline{\sl and Istituto Nazionale di Fisica Nucleare, Gruppo collegato
di Cosenza}
\centerline{\sl I-87036 Arcavacata di Rende, Cosenza, Italy}
\vskip 0.6cm

\begin{abstract}
Critical properties of the compact three-dimensional $U(1)$ lattice gauge 
theory are explored at finite temperatures on an asymmetric lattice. For 
vanishing value of the spatial gauge coupling one obtains an effective 
two-dimensional spin model which describes the interaction between Polyakov 
loops. We study numerically the effective spin model for $N_t=1,4,8$ on 
lattices with spatial extension ranging from $L=64$ to $L=256$. Our results 
indicate that the finite-temperature $U(1)$ lattice gauge theory belongs to 
the universality class of the two-dimensional $XY$ model, thus supporting the 
Svetitsky-Yaffe conjecture.
\end{abstract}

\vfill
\hrule
\vspace{0.3cm}
{\it e-mail addresses}: 
\noindent$^*$oleg@bitp.kiev.ua, 
\noindent$^{**}$gravina,papa@cs.infn.it 

\newpage

\section{Introduction}

The finite temperature behavior of the compact three-dimensional ($3d$) $U(1)$ 
lattice gauge theory (LGT) is the subject of numerous investigations (see, 
e.g., Ref.~\cite{chernodub} and references therein). 
It is well-known that at zero temperature the theory is confining at all values
of the bare coupling constant~\cite{polyakov}. 
At finite temperature the theory undergoes a deconfinement 
phase transition. Both phenomena are expected to take place in $4d$ QCD as 
well. Therefore, the $3d$ $U(1)$ gauge theory constitutes one of the simplest 
models with continuous gauge symmetry which possess the same fundamental 
properties as QCD. In view of these common features the critical properties of 
$3d$ $U(1)$ LGT deserve comprehensive qualitative and quantitative 
understanding.

On the theoretical side one should mention two results regarding the critical 
behavior of $3d$ $U(1)$ LGT. The first result states that the partition 
function of $3d$ $U(1)$ LGT in the Villain formulation coincides with that of 
the $2d$ $XY$ model in the leading order of the high-temperature 
expansion~\cite{parga}. In particular, the monopoles of the original $U(1)$ 
gauge theory are reduced to vortices of the $2d$ system. The second result 
follows from the Svetitsky-Yaffe conjecture: the finite-temperature phase 
transition in the $3d$ $U(1)$ LGT should belong to the universality class of 
the $2d$ $XY$ model if correlation length diverges~\cite{svetitsky}. Then,
two possibilities arise: either the transition is first order or it
is the same transition which occurs in the $2d$ $XY$ model.
The $XY$ model is known to have Berezinskii-Kosterlitz-Thouless (BKT) 
phase transition of infinite order~\cite{Berezinsky:1970fr,Kosterlitz:1973xp}.
Several important facts could be deduced from these results.
First of all, the global $U(1)$ symmetry cannot be broken spontaneously even at
high temperatures because of the Mermin-Wagner theorem. Consequently, a local 
order parameter does not exist. Secondly, one might expect the critical 
behavior of the Polyakov loop correlation function $\Gamma (R)$ to be governed 
by the following expressions
\begin{equation}
\Gamma (R) \ \asymp \ \frac{1}{R^{\eta (T)}} \ ,
\label{PLhight}
\end{equation}
for $\beta \geq \beta_c$ and 
\begin{equation}
\Gamma (R) \ \asymp \ \exp \left [ -R/\xi (t)  \right ] \ ,
\label{PLlowt}
\end{equation}
for $\beta < \beta_c$, $t=\beta_c/\beta -1$.
Here, $R\gg 1$ is the distance between test charges and $\xi \sim 
e^{bt^{-\nu}}$ is the correlation length. Such behavior of $\xi$ defines the 
so-called {\it essential scaling}. The critical indices $\eta (T)$ and $\nu$ 
are known from the renormalization-group analysis of the $XY$ model: 
$\eta (T_c) =1/4$ and $\nu=1/2$, where $T_c$ is the BKT critical point. 
Therefore, the critical indices $\eta$ and $\nu$ should be the same in the 
finite-temperature $U(1)$ model if the Svetitsky-Yaffe conjecture holds in 
this case.

The first renormalization-group calculations of the critical indices, 
presented in~\cite{svetitsky}, gave support to the conjecture even though 
they did not constitute a rigorous proof. The direct numerical check of 
these predictions was performed on lattices $N_s^2\times N_t$ with $N_s=16, 
32$ and $N_t=4,6,8$ in Ref.~\cite{mcfinitet}. Though authors of 
Ref.~\cite{mcfinitet} confirm the expected BKT nature of the phase transition, 
the reported critical index is almost three times that predicted for the $XY$ 
model, $\eta (T_c) \approx 0.78$. More recent numerical simulations of 
Ref.~\cite{chernodub} have been mostly concentrated on the study of the 
properties of the high-temperature phase. We have to conclude that, so far, 
there are no numerical indications that the critical indices of $3d$ $U(1)$ 
LGT do coincide with those of the $2d$ $XY$ model. Moreover, since a rigorous 
determination of the critical indices is not available even for the $XY$ model 
one can hardly hope for a rigorous analysis of the critical behavior of $3d$ 
$U(1)$ LGT.

The absence of reliable results in the vicinity of the BKT critical point was 
our primary motivation to study the deconfinement phase transition in $3d$ 
$U(1)$ LGT. The difficulties in computations of critical indices of the $XY$ 
model are well-known and we do not intend to discuss them here (see 
Ref.~\cite{kenna} for a summary of recent results and problems). It should be 
clear however, that in the context of the $3d$ theory a reliable determination 
of critical properties becomes even harder and requires simulations on very 
large lattices. We have decided therefore to attack the problem in a few steps.
Consider the finite-temperature model on anisotropic lattice with different 
spatial and temporal coupling constants; as a first step, in this paper 
we investigate the limit of vanishing spatial coupling. The major advantage of 
this limit is that the integration over spatial links can be performed 
analytically. The result of such integration is an effective two-dimensional 
spin model for the Polyakov loops. The latter can be studied numerically.   

This paper is organized as follows. In the next section we introduce the 
compact $U(1)$ LGT on anisotropic lattice and study it for vanishing spatial 
coupling. In the Section 3 we describe briefly our numerical procedure. The 
result of simulations are presented in the Section 4. Conclusions and 
perspectives are given in the Section 5.

\section{The 3$d$ U(1) lattice gauge theory}
\label{theory}

We work on a $3d$ lattice $\Lambda = L^2\times N_t$ with spatial extension $L$ 
and temporal extension $N_t$. Periodic boundary conditions on gauge fields are 
imposed in all directions. We introduce anisotropic dimensionless couplings in 
a standard way as 
\begin{equation}
 \beta_t = \frac{1}{g^2a_t}  \ , \;\;\;\;\; \beta_s = \frac{\xi}{g^2a_s} \ = \ 
\beta_t \ \xi^2 \ , \;\;\;\;\; \xi = \frac{a_t}{a_s} \ ,
\label{ancoupl}
\end{equation}
where $a_t$ ($a_s$) is lattice spacing in the time (space) direction.
$g^2$ is the continuum coupling constant with dimension $a^{-1}$.
The finite-temperature limit is constructed as 
\begin{equation}
\xi\to 0 \ , \;\;\;\;\; \ N_t \ , \ L \ \to\infty \ , \;\;\;\;\; \ a_tN_t = 
\frac{1}{T} \ ,
\label{fintemplim}
\end{equation}
where $T$ is the temperature.

The $3d$ $U(1)$ LGT on the anisotropic lattice is defined through its partition
function as 
\begin{equation}
Z(\beta_t,\beta_s) = \int_0^{2\pi}\prod_{x\in\Lambda}\: \prod_{n=0}^2
\frac{d\omega_n (x)}{2\pi} \ \exp{S[\omega]} \ ,
\label{PTdef}
\end{equation}
where $S$ is the Wilson action 
\begin{equation}
 S[\omega] = \beta_s\sum_{p_s} \cos\omega (p_s) + 
\beta_t\sum_{p_t} \cos\omega (p_t) 
\label{wilsonaction} 
\end{equation}
and sums run over all space-like ($p_s$) and time-like ($p_t$) plaquettes.
The plaquette angles $\omega(p)$ are defined in the standard way.
The correlation of two Polyakov loops can be written as, e.g.
\begin{equation}
\Gamma (R) \ = \  \left \langle  
\exp \left [ i\sum_{x_0=0}^{N_t-1}(\omega_0(x_0,x_1,x_2) -
\omega_0(x_0,x_1,x_2+R)) \right ] \right \rangle \ \ .
\label{PL}
\end{equation}

As stated in the Introduction we would like to explore the limit $\beta_s =0$. 
Consider the strong coupling expansion at $\beta_s \ll 1$. 
The general form of such expansion reads
\begin{equation}
Z(\beta_t,\beta_s) = 
Z(\beta_t,\beta_s=0) + \sum_{k=1} \beta_s^{2k} \ Z_{2k}(\beta_t) \ .
\label{PTexp}
\end{equation}
In this paper we study the zero-order partition function $Z(\beta_t,\beta_s=0)$
defined below. The series on the right-hand side of the last expression is 
known to be convergent uniformly in the volume, both for the free energy and 
for the gauge-invariant correlation functions. The uniform convergence 
guarantees the existence of the limit $N_t\to\infty$. The strong coupling 
expansion, done even in one parameter, might be far from the continuum limit. 
Nevertheless, one expects that already the zero-order approximation captures 
correctly the critical behavior of the full theory. An example is given by 
the following Polyakov loop model
\begin{equation}
S_{\mbox{\scriptsize eff}} = \beta_{\mbox{\scriptsize eff}}
\sum_{x,n}{\rm Re} \  W(x)W^*(x+n) \ ,
\label{PLM}
\end{equation}
derived at finite temperature for $(d+1)$-dimensional $SU(N)$ pure gauge theory
in the limit $\beta_s=0$. Here, $\beta_{\mbox{\scriptsize eff}}\propto 
\beta_t^{N_t}$. As is well known, this model reveals correctly the critical 
behavior of the original theory thus supporting our approximation.

In the zero-order approximation the integration over spatial gauge fields can 
be easily done and leads to the following expression for the partition function
\begin{equation}
 Z(\beta_t,\beta_s=0) \ = \ \int_0^{2\pi}\prod_{x}\frac{d\omega_x}{2\pi} 
\prod_{x,n} \left [ \sum_{r=-\infty}^{\infty} \  
I_r^{N_t}(\beta_t) \ \exp \left [ ir(\omega (x)-\omega (x+e_n)) \right ] 
\right ] \ , 
\label{pfbs0pl}
\end{equation}
where $x$ belongs to the two-dimensional lattice $\Lambda_2=L^2$ and $\omega(x)
\equiv \omega(x_1,x_2)$. Here, $I_r(x)$ are modified Bessel functions and
$e^{ir\omega(x)}$ is the Polyakov loop in the representation $r$. 

For $N_t=1$ using the formula $\sum_rI_r(x)e^{ir\omega}=e^{x\cos\omega}$ one 
finds 
\begin{equation}
\left. Z(\beta_t,\beta_s=0)\right|_{N_t=1} \ = \ \int_0^{2\pi}\prod_{x}
\frac{d\omega(x)}{2\pi} 
\ \exp \left [ \beta_t \sum_{x,n} \cos (\omega (x) - \omega (x+e_n)) \right ] \
\label{pfbs0nt1}
\end{equation}
which is the partition function of the $2d$ $XY$ model. Thus, in this case the 
dynamics of the system is governed by the $XY$ model with the inverse 
temperature $\beta_t$. 
For $N_t\geq 2$ the model (\ref{pfbs0pl}) is of the $XY$-type, i.e. 
it describes interaction between nearest neighbors spins (Polyakov loops) and 
possesses the global $U(1)$ symmetry.  Moreover, consider now two different 
limits - the strong coupling limit $\beta_t\ll 1$ and the weak coupling limit 
$\beta_t\gg 1$.

In the leading order of the strong coupling limit one can easily find from 
(\ref{pfbs0pl}), up to an irrelevant constant,
\begin{equation}
 Z(\beta_t\ll 1,\beta_s=0) \ = \ \int_0^{2\pi}\prod_{x}\frac{d\omega(x)}{2\pi} 
\ \exp \left [ h(\beta_t) \sum_{x,n} \cos (\omega (x) - \omega (x+e_n)) 
\right ] \  
\label{pfbs0btl1}
\end{equation}
which is again the $XY$ model with the coupling $h$ given by 
$$
h(\beta_t) \ = \ 2 \left [ \frac{I_1({\beta_t})}{I_0(\beta_t)} \right ]^{N_t} 
\ .
$$
The Polyakov loop vanishes while the correlations of the Polyakov loops are 
given, at the leading order, by
\begin{equation}
 \Gamma (R) \ = \  \left [ \frac{1}{2} \ h(\beta_t) \right ]^R \ .
\label{PLbs0}
\end{equation}

To study the weak coupling limit it is convenient to perform duality 
transformations which are well-known for the $XY$ model. Taking then the 
asymptotics of the Bessel functions one obtains, up to an irrelevant constant,
\begin{equation}
 Z(\beta_t\gg 1,\beta_s=0) \ = \ \sum_{r(x)=-\infty}^{\infty} \  
 \exp{ \left [ - \frac{1}{2}\tilde{\beta} 
 \sum_x \sum_{n=1}^{2} (r(x)-r(x+e_n))^2 \right ] } \ . 
\label{pfbs0btgg1}
\end{equation}
This is nothing but the Villain version of the $XY$ model in the dual 
formulation with an effective coupling
\begin{equation}
 \tilde{\beta} \ = \ N_t/\beta_t \ = \ g^2/T \ .
\label{betatild}
\end{equation}
This shows that the region $\beta_s=0$, $\beta_t\gg 1$ is also described by 
the $XY$ model. 

In the general case of arbitrary $\beta_t$ the full effective action 
\begin{equation}
 S_{\mbox{\scriptsize eff}} \ = \ \sum_{x,n} \sum_k \ C_k 
\cos k(\omega (x) - \omega (x+e_n))
\label{effaction}
\end{equation}
will include all representations $k$ of the Polyakov loops. In our case 
the coefficients $C_k$ are given by
\begin{equation}
 C_k \ = \ \int_0^{2\pi}\frac{d\omega}{2\pi} \ \cos k\omega  \
 \log \biggl\{ 1+ 2 \sum_{r=1}^\infty [b_r(\beta_t)]^{N_t} \cos r\omega  
\biggr\} \ ,
\label{Ck}
\end{equation}
where $b_r(\beta_t)=I_r(\beta_t)/I_0(\beta_t)$. If there is a critical point
at which the correlation length is divergent then on general grounds 
(universality, limiting behavior, etc.) one assumes that the model described 
by the effective action~(\ref{effaction}) indeed possesses the same critical 
behavior as the $XY$ model. Nevertheless, we are not aware of any direct 
numerical check of the universality for models of the type~(\ref{effaction}) if
$C_k \ne 0$ for $k=2,3,...$. In the following sections we present numerical 
simulations which give support for the expected BKT behavior of the model 
(\ref{effaction}). Our results hold only for the model with $C_k$ defined 
by (\ref{Ck}). We would like to stress that it is not obvious that for all 
possible $C_k$ the correlation length really diverges. For example, it was 
proven in Ref.~\cite{ES02} that the model with coefficients
\begin{equation}
C_k \ = \ \int_0^{2\pi}\frac{d\omega}{2\pi} \ \cos k\omega  \
\left(\frac{1+ \cos \omega}{2} \right)^p \ ,
\end{equation}
with sufficiently large $p$, exhibits a first order phase transition, so that 
one could expect that the correlation length stays finite across the phase 
transition point.

\section{Numerical set-up}
\label{setup}

Determining the universality class of the 3$d$ U(1) gauge theory discretized
on a $L^2 \times N_t$ lattice means determining its critical indices. 
A convenient way to accomplish this task is to study the scaling with the 
spatial size $L$ of the vacuum expectation value of suitable observables, 
determined through numerical Monte Carlo simulations. 

For the special case $\beta_s=0$, one can take advantage of Eq.~(\ref{pfbs0pl})
and describe the original gauge system with a two-dimensional spin model whose 
action $S'$ is defined through
\be
Z(\beta_t,\beta_s=0) \equiv \int_0^{2\pi}\prod_x \frac{d\omega(x)}{2\pi} 
\exp S'
\label{spin_part} 
\ee
and reads
\begin{equation}
S'= \sum_{x,n} \log \biggl\{ 1+ 2 \sum_{r=1}^\infty [b_r(\beta_t)]^{N_t} 
\cos r(\omega (x) - \omega (x+e_n)) \biggr\}\;.
\label{spin_act}
\end{equation}
The infinite series in $r$ can be truncated early, since the $b_r$'s vanish 
very rapidly for increasing $r$. We studied the dimensionally reduced system 
with the Metropolis algorithm, taking the first twenty $b_r$ couplings
(notice that $b_{20}(\beta_t=1) \sim 10^{-25}$).
 
Our goal is to bring evidence that the system exhibits BKT critical 
behavior {\em for any fixed $N_t$}. This is trivially verified in the case 
$N_t=1$, since by inspection of Eqs.~(\ref{spin_part}) and~(\ref{spin_act}), 
the theory reduces exactly to the $XY$ model. Therefore the case $N_t=1$ can 
be used as a test-field for the description and the validation of our 
procedure.

Before presenting numerical results it is instructive to give some
simple analytical predictions for the critical values $\beta_t$ at
different  values of $N_t$. Such critical values can be easily estimated
if one knows $\beta_t^{cr}$ for $N_t=1$. Since the model with $N_t=1$
coincides with the $XY$ model one has $\beta_t^{cr}(N_t=1)\approx 1.119$
and approximate critical points for other values of $N_t$ can be computed
from the equality
\begin{equation}
b_1(1.119) \ = \ [b_1(\beta_t^{cr})]^{N_t} \ .
\label{crpointest}
\end{equation}
Solving the last equation numerically one finds $\beta_t^{cr}$. The results
are given in the Table~\ref{beta_crest}. As will be seen below the predicted
values are in a reasonable agreement with the numerical results.

\begin{table}[ht]
\centering\caption[]{Analytical estimates of $\beta_t^{cr}$ for several
values of $N_t$ (first row) compared with the numerical results obtained
in Section~\ref{results} (second row).}
\vspace{0.2cm}
\begin{tabular}{|c|c|c|c|c|}
\hline
 $N_t$ & $2$ & $4$ & $8$ & $16$  \\
\hline
$\beta_t^{cr}$ & 2.0003 & 3.39389 & 6.10642 & 11.6385 \\
               &        & 3.42(1) & 6.38(5) &         \\ 
\hline
\end{tabular}
\label{beta_crest}
\end{table}

\section{Results at $\beta_s=0$}
\label{results}

\subsection{$N_t=1$}

The main indication of BKT critical behavior is a peculiar
scaling of the pseudo-critical coupling with the spatial lattice size $L$,
consequence of the {\em essential scaling},~\footnote{Throughout this Section
we use the notation $\beta_t\equiv \beta$.} 
\be
\beta_{pc}(L) - \beta_c \sim \frac{1}{(\log L)^{1/\nu}}\;,
\label{beta_scaling}
\ee
where $\beta_{pc}(L)$ is the pseudo-critical coupling on a lattice
with spatial extent $L$, $\beta_c$ is the (non-universal) infinite volume 
critical coupling and $\nu$ is the (universal) thermal critical index.

\begin{figure}[tb]
\centering
\includegraphics[width=13cm]{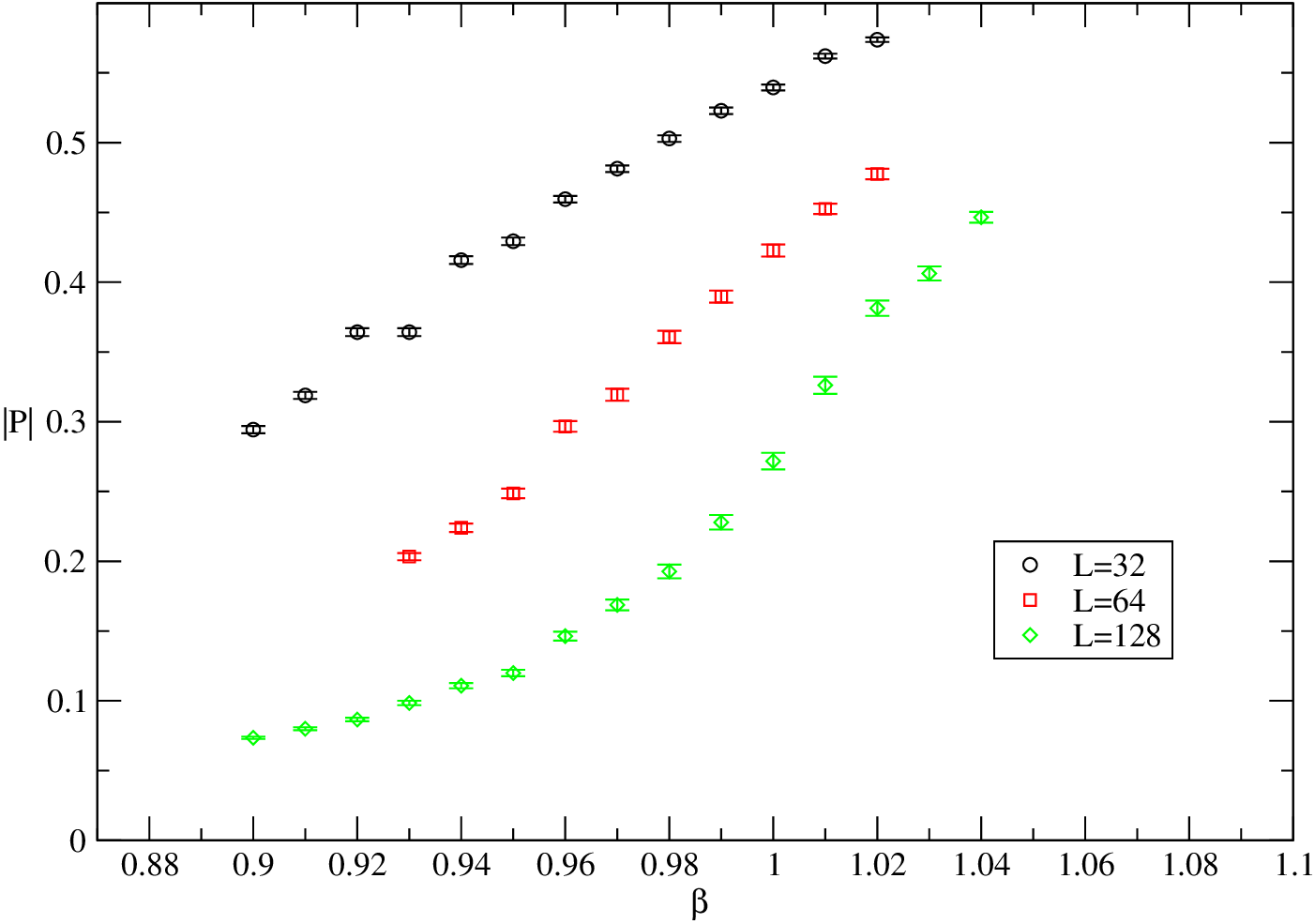}

\vspace{1.5cm}
\includegraphics[width=13cm]{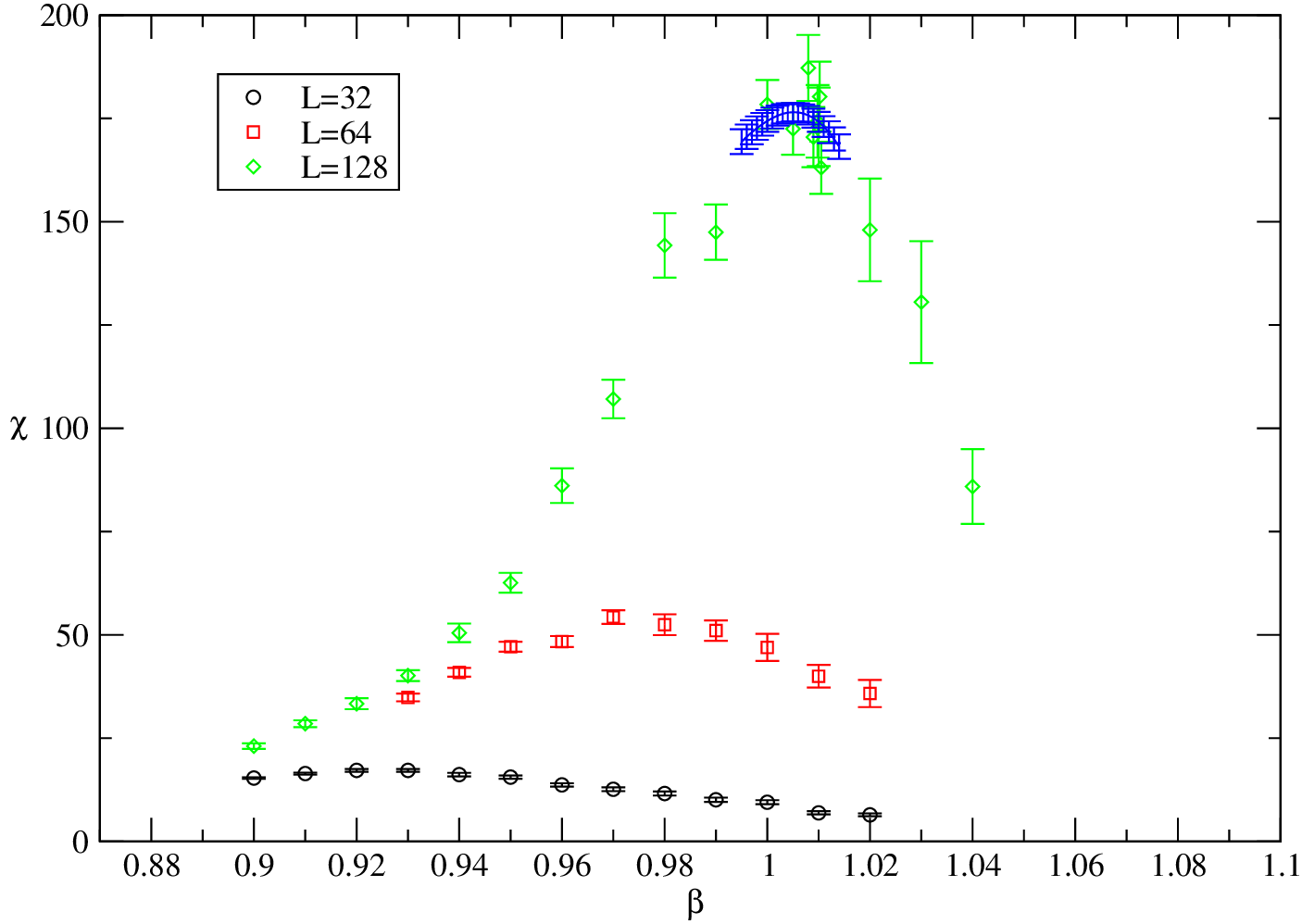}
\caption{(Top) Absolute value of the Polyakov loop {\it vs} $\beta$
on a $1\times L^2$ lattice, with $L$=32, 64, 128. 
(Bottom) Susceptibility of the Polyakov loop {\it vs} $\beta$ 
on a $1\times L^2$ lattice, with $L$=32, 64, 128. For the $L$=128 case
the multi-histogram interpolation around the peak is shown.}
\label{poly_chi}
\end{figure}

The pseudo-critical coupling $\beta_{pc}(L)$ is determined by the value
of $\beta$ for which a peak shows up in the susceptibility of the Polyakov 
loop,
\be
\chi = L^2 \langle |P|^2 \rangle \; ,
\;\;\;\;\;\; P = \frac{1}{L^2} \sum_x P_x\;;
\label{poly_chi_def} 
\ee
here the local Polyakov loop variable $P_x$ corresponds to the spin 
$s_x=\exp{i\omega(x)}$ of the $XY$ model.
In Fig.~\ref{poly_chi} we show the behavior of the absolute value
of the Polyakov loop $|P|$ (top) and of the susceptibility $\chi$ (bottom),
for varying $\beta$ on lattices with $L$= 32, 64, 128.

To extract $\beta_{pc}(L)$ in a more reliable way, we performed the 
multi-histogram interpolation~\cite{multihisto}; errors were determined
by the jackknife method. Results for $\beta_{pc}(L)$ are summarized in 
Table~\ref{beta_pc}.

\begin{table}[ht]
\centering\caption[]{$\beta_{pc}(L)$ for $N_t$=1, 4, 8 and for several   
values of $L$. Errors are determined by a jackknife analysis.}
\vspace{0.2cm}
\begin{tabular}{|c|c|c|c|}
\hline
 $L$ & $N_t=1$ & $N_t=4$ & $N_t=8$  \\
\hline
64& - & 3.1250(51) & 5.531(19) \\
128& 1.0051(16)& - & 5.754(22) \\
150& 1.0094(26)& 3.2190(40) & 5.7945(59) \\
200& 1.0227(15)& 3.2368(39) & - \\
256& 1.0278(20)& - & - \\
\hline
\end{tabular}
\label{beta_pc}
\end{table}

We determined $\beta_c(N_t=1)$ by fitting the pseudo-critical coupling
$\beta_{pc}(L)$ given in the second column of Table~\ref{beta_pc} with
the law 
\be
\beta_{pc}(L)=\beta_c+\frac{A}{(\log L)^{1/\nu}}\;,
\label{essential}
\ee
in which $\nu$ was fixed by hand at the $XY$ value, $\nu=1/2$. We got
$\beta_c(N_t=1)=1.107(9)$ and $A(N_t=1)=-2.4(2)$ ($\chi^2$/d.o.f.=0.78), 
which is quite in agreement with the best known $XY$ critical 
coupling, $\beta_c=1.1199(1)$, given in Ref.~\cite{Hasenbusch-Pinn}.

The determination of $\beta_c$ is crucial in order to extract critical indices;
indeed, they enter scaling laws which hold just at $\beta_c$, such as, for 
example, 
\be
\chi(\beta_c) \sim L^{2-\eta_c} \quad ,
\label{chi_scaling}
\ee
where $\eta_c$ is the magnetic critical index. Actually in 
Eq.~\ref{chi_scaling} one should consider logarithmic corrections 
(see~\cite{Kenna-Irving,Hasenbusch} and references therein) and, indeed,
recent works on the $XY$ universality class generally include them. However,
taking these corrections into account for extracting critical indices calls
for very large lattices even in the $XY$ model; for the theory under 
consideration to be computationally tractable, we have no choice but to neglect
logarithmic corrections.

We determined $\chi(\beta=1.12)$ for $L$=64, 128, 150, 200, 256 -- see
Table~\ref{chi_L} for a summary of the results. Fitting with the 
law~(\ref{chi_scaling}), we found $\eta_c=0.256(29)$ ($\chi^2$/d.o.f.=0.2), 
in nice agreement with the $XY$ value, $\eta_c=1/4$. The same analysis repeated
at $\beta=1.107$, {\it i.e.} at the central value of our determination
of $\beta_c$, on lattices with $L$=64, 128, 200, gave $\eta_c=0.237(61)$ 
($\chi^2$/d.o.f.=0.01).

\begin{figure}[tb]
\centering

\includegraphics[width=13cm]{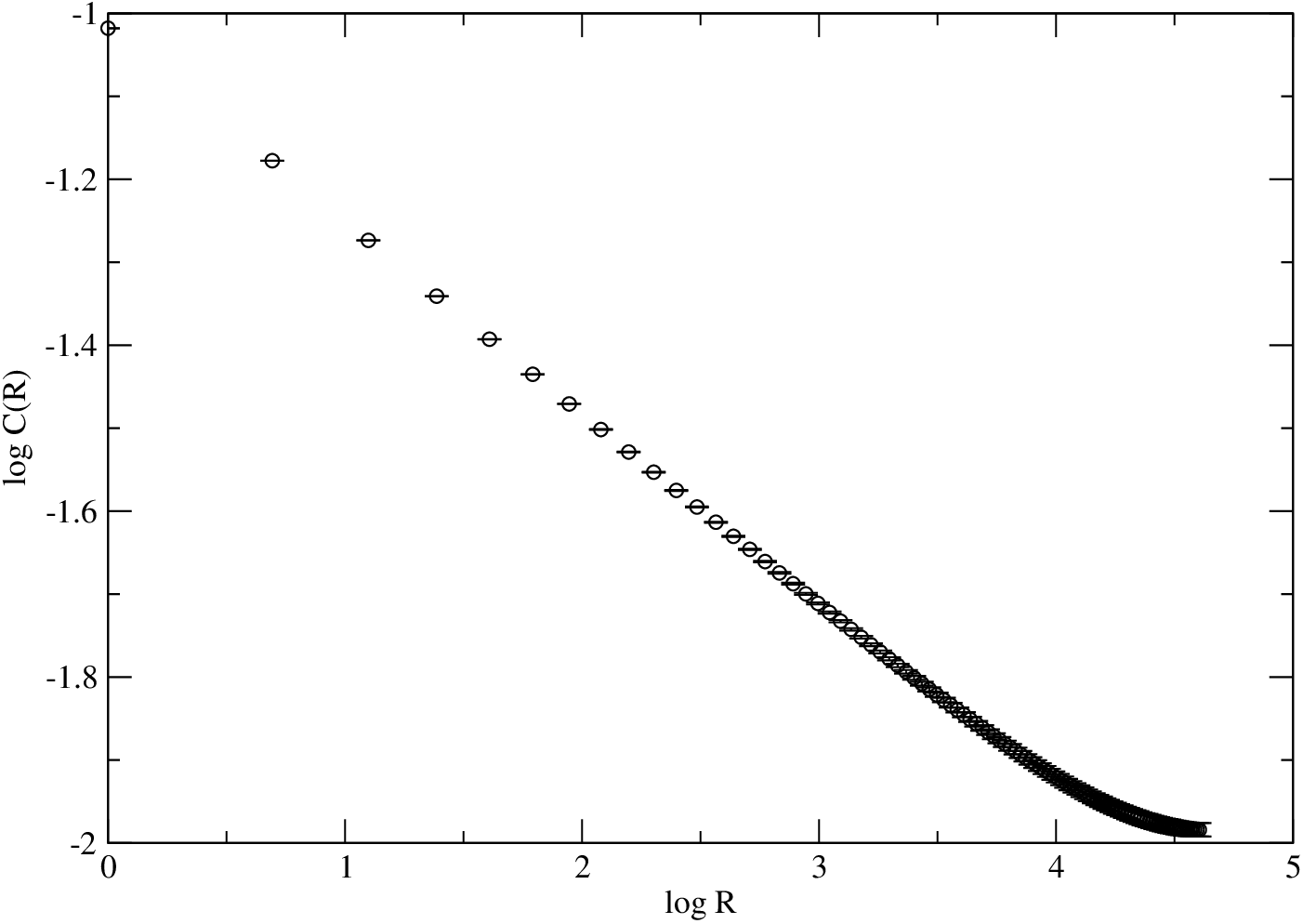}
\caption[]{Log-log plot of point-point correlator for $L$=200 at 
$\beta=1.12$.}
\label{log_corr}
\end{figure}

An alternative strategy to determine $\eta_c$ uses the large 
distance behavior of the point-point correlator of the Polyakov loop, 
\be
C(R) = \sum_{x,n} \Re{ \bigg( P_x^\dagger  P_{x + R \cdot e_n} \bigg)} \quad ,
\label{pp_corr1}
\ee
where $e_n$ is the unit vector in the $n$-th direction. Without logarithmic    
corrections, one has
\be
C(R) \sim \frac{1}{R^{\eta_c}} \quad .
\label{pp_corr2}
\ee
In Fig.~\ref{log_corr} we plot $\log C(R)$ {\it vs} $\log R$ for $L$=200
at $\beta=1.12$; linearity is clear up to $R \simeq 30$. Deviations at larger 
distances are due to finite size effects (echo terms are expected to be 
strong, since the correlator is long-ranged) and possibly to logarithmic    
corrections. In the linear regime ($5 < R < 30$), the naive fit with a 
power law gives $\eta$= 0.22942(31) ($\chi^2$/d.o.f.=0.83).
The same analysis at $\beta=1.107$ and $L$=200 gives $\eta$= 0.2380(20) 
($\chi^2$/d.o.f.=0.05) in the range $1 < R < 45$. On the same volume
one sees that, for lower $\beta$'s, $\eta$ goes towards the expected value 
and that the linear region gets wider and wider.

\begin{figure}[tb]
\centering

\includegraphics[width=13cm]{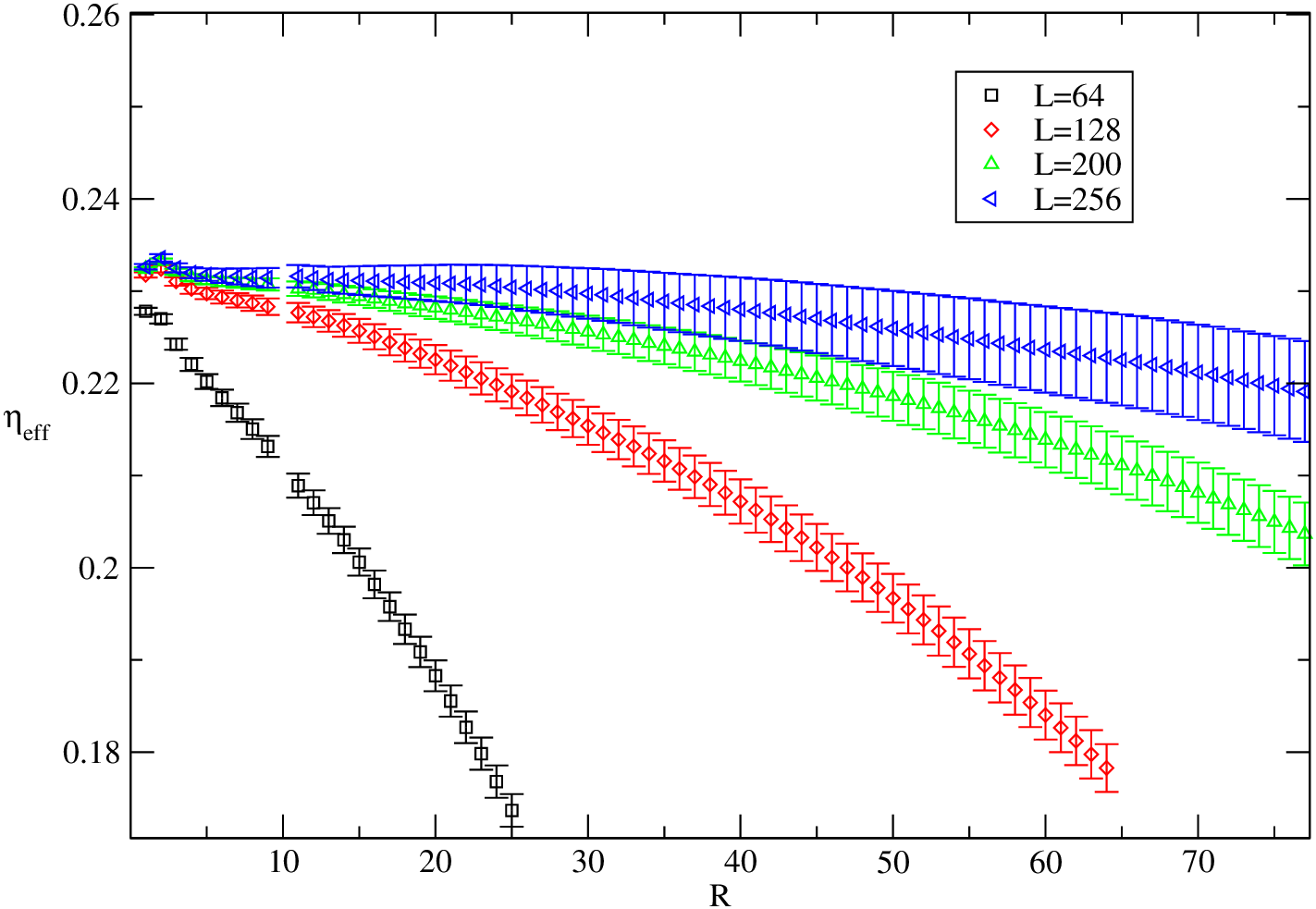}

\vspace{1.5cm}
\includegraphics[width=13cm]{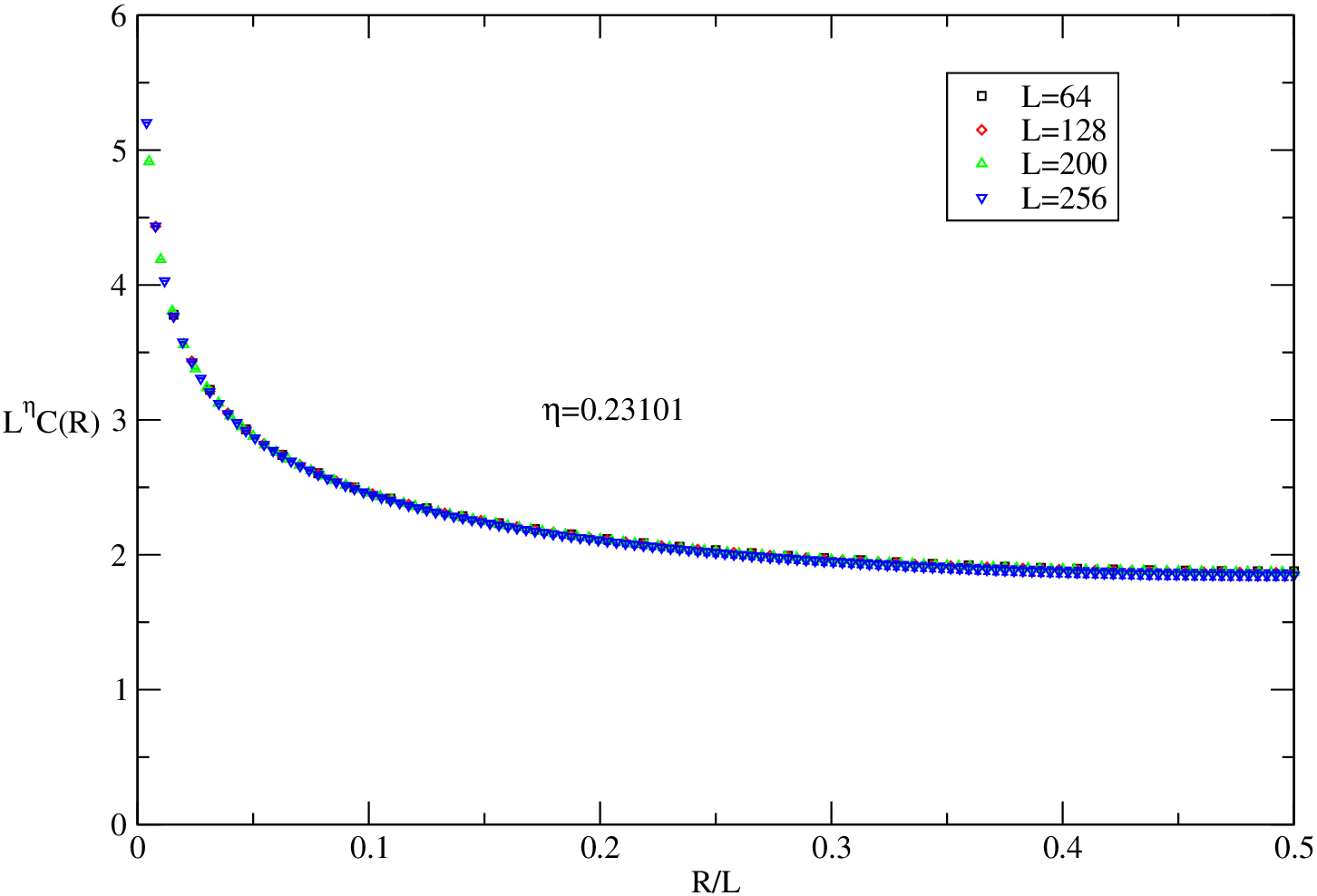}
\caption[]{(Top) $\eta_{\mbox{\scriptsize eff}}$ for $N_t=1$ on lattices 
with several spatial sizes $L$ at $\beta=1.12$. For all lattices we fixed 
$R_0$=10. Errors are calculated with the jackknife method. 
(Bottom) $L^\eta C(R)$ versus $R/L$, with $\eta$ fixed at the central value 
of our determination through the method of the effective 
$\eta_{\mbox{\scriptsize eff}}$ (see the text).}

\label{eta_1}
\end{figure}

The {\it effective $\eta_c$} index, defined as
\be
\eta_{\mbox{\scriptsize eff}}(R) \equiv \frac{\log [C(R)/C(R_0)]}
{\log [R_0/R]} \quad ,
\label{eff_eta}
\ee
must exhibit a plateau in the region where~(\ref{pp_corr2}) holds.
Fig.~\ref{eta_1}(top) shows that the larger the volume the larger
the region in which there is a plateau at small distances. The chosen
value of $R_0$ must belong to the linear region in order to minimize finite 
size effects. We have verified that varying $R_0$ in the linear region does 
not change the result and have chosen $R_0=10$ for all the cases considered 
here.

Since for the larger lattices ($L$=200 and $L$=256) plateaux are overlapping 
at small distances, one can conclude that thermodynamic limit is reached.
We estimate the plateau value from the most precise data we have ($L$=200) as
$\eta(\beta=1.12)=\eta_{\mbox{\scriptsize eff}}(R=6)=0.23101(49)$, 
since the latter is the value of 
$\eta_{\mbox{\scriptsize eff}}$ in the linear region compatible with
the largest number of subsequent points. Deviations from the expected value 
$\eta$=0.25 can be due either to logarithmic corrections or to the 
overestimation of $\beta_c$. Repeating the same procedure for slightly 
lower $\beta$'s we find: $\eta(\beta=1.115)=0.23491(47)$ and 
$\eta(\beta=1.107)=0.24085(44)$. Notice that $\eta$ approaches the expected 
value as $\beta$ lowers. The relation between $\eta$ and $\beta$ is well   
described by a linear function ($\chi^2$/d.o.f.=0.04) and this suggests that 
the $\beta$ value at which $\eta$=0.25 is really close to those considered.
Fig.~\ref{eta_1}(bottom) shows the correlation function $C(R)$ rescaled
by $L^{-\eta}$ in units of $R/L$; it turns out that, when the best 
determination for $\eta$ is used (in the present case, $\eta=0.23101$)
data from different lattices fall on top of each other over almost all
the range of distances considered. 

There are other observables which turned out to be useful in establishing
the BKT scaling in the $2d$ $XY$ model and which we do not use in the present
work: the helicity modulus ${\cal Y}$~\cite{WM88,Hasenbusch}, the second
moment correlation length $\xi_2$ (see, for instance,~\cite{Hasenbusch})
and the $U_4$ cumulant, proposed in~\cite{Has08}. We plan to use them all when 
we will study the general case $\beta_s\neq 0$. For the purposes of the
present work we have only tried to use the $U_4$ cumulant, but both lattice 
sizes and statistics seem to be not enough large to extract any useful 
information from this observable.
 
\subsection{$N_t$=4 and 8}

In this Subsection we extend the study performed in the $N_t=1$ case to
the cases of $N_t=4$ and 8, with the aim of showing that the universal    
$XY$ features are not lost increasing $N_t$ at $\beta_s$=0.

In Table~\ref{beta_pc} we give the values of the pseudo-critical 
couplings $\beta_{pc}(L)$ obtained from the peaks of the Polyakov
loop susceptibility for several values of $L$ at $N_t = 4$ and 8.
Fitting these values with the law~(\ref{essential}) with $\nu$ = 1/2
fixed, we get
\begin{eqnarray}
\beta_c(N_t=4)&=&3.42(1)\;, \;\;\;\;\; A(N_t=4)=-5.1(3)\;, \;\;\;\;\;
(\chi^2/\mbox{d.o.f.}=0.43) \nonumber 	\\
\beta_c(N_t=8)&=&6.38(5)\;, \;\;\;\;\; A(N_t=8)=-15(1)\;, \;\;\;\;\;
(\chi^2/\mbox{d.o.f.}=0.006) \quad . 
\label{bc_4_8}
\end{eqnarray}
This result shows that essential scaling is satisfied, {\it i.e.}
in both cases transition is compatible with BKT. It is worth noting that 
these values of $\beta_c$ are in nice agreement with the estimates given in
Table~\ref{beta_crest}. This suggests that the dynamics of the effective
model near the transition point is indeed dominated by the lower 
representations, thus justifying the truncation of the series in 
Eq.~(\ref{spin_act}). 

\begin{table}[ht]
\centering\caption[]{$\chi(L)$ for $N_t$=1, 4, 8. Errors are determined by
a jackknife analysis.}
\vskip 0.4cm
\begin{tabular}{|c|c|c|c|}
\hline
 $L$ & $N_t=1$ & $N_t=4$ & $N_t=8$  \\
\hline
64  &  7.19(12) & 9.30(57)    & 7.33(37) \\
128 & 24.6(1.3)  & 35.9(4.4)  & 25.1(1.5) \\
150 & 32.9(1.9)  & 42.5(2.5)  & 32.5(1.9) \\
200 & 51.4(2.7)  & 65.2(2.8)  & 58.4(3.3) \\
256 & 80.3(4.2)  & 101.7(5.4) & 86.4(3.6) \\
\hline
\end{tabular}
\label{chi_L}
\end{table}

In Table~\ref{chi_L} we give the values of the Polyakov loop susceptibility 
for several values of $L$ at $\beta=3.42$ for $N_t=4$ and at $\beta=6.38$
for $N_t=8$. Fitting with~(\ref{chi_scaling}), we find
\begin{eqnarray}
\eta_c(N_t=4)&=&0.290(54) \quad (\chi^2/\mbox{d.o.f.}=0.69) \nonumber \\
\eta_c(N_t=8)&=&0.212(46) \quad (\chi^2/\mbox{d.o.f.}=0.28) \quad . 
\label{eta_4_8}
\end{eqnarray}
Results agree with the universal $XY$ value $\eta_c=1/4$, although errors are 
quite large. 

\begin{figure}[tb]
\centering
\includegraphics[width=13cm]{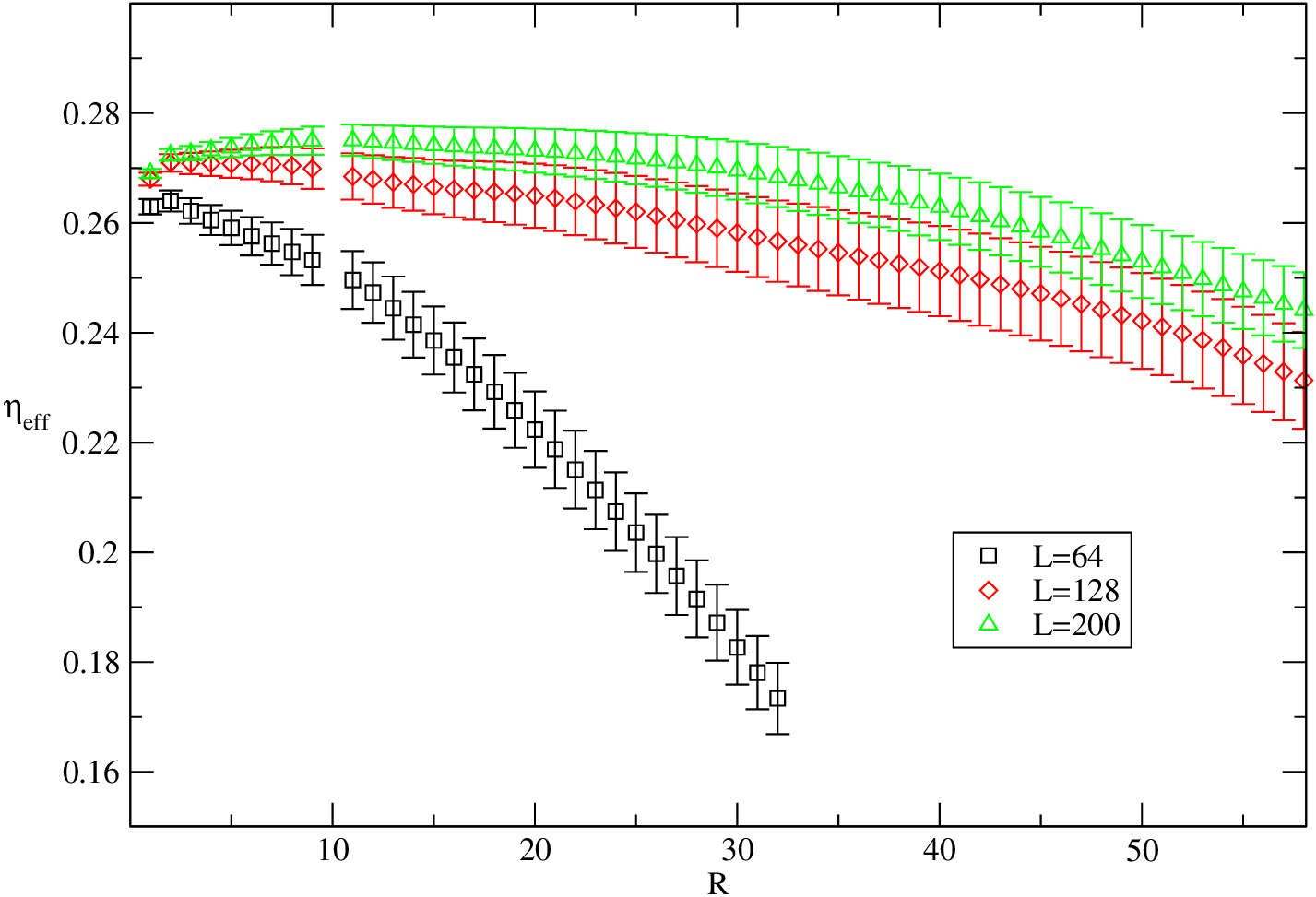}

\vspace{1.5cm}
\includegraphics[width=13cm]{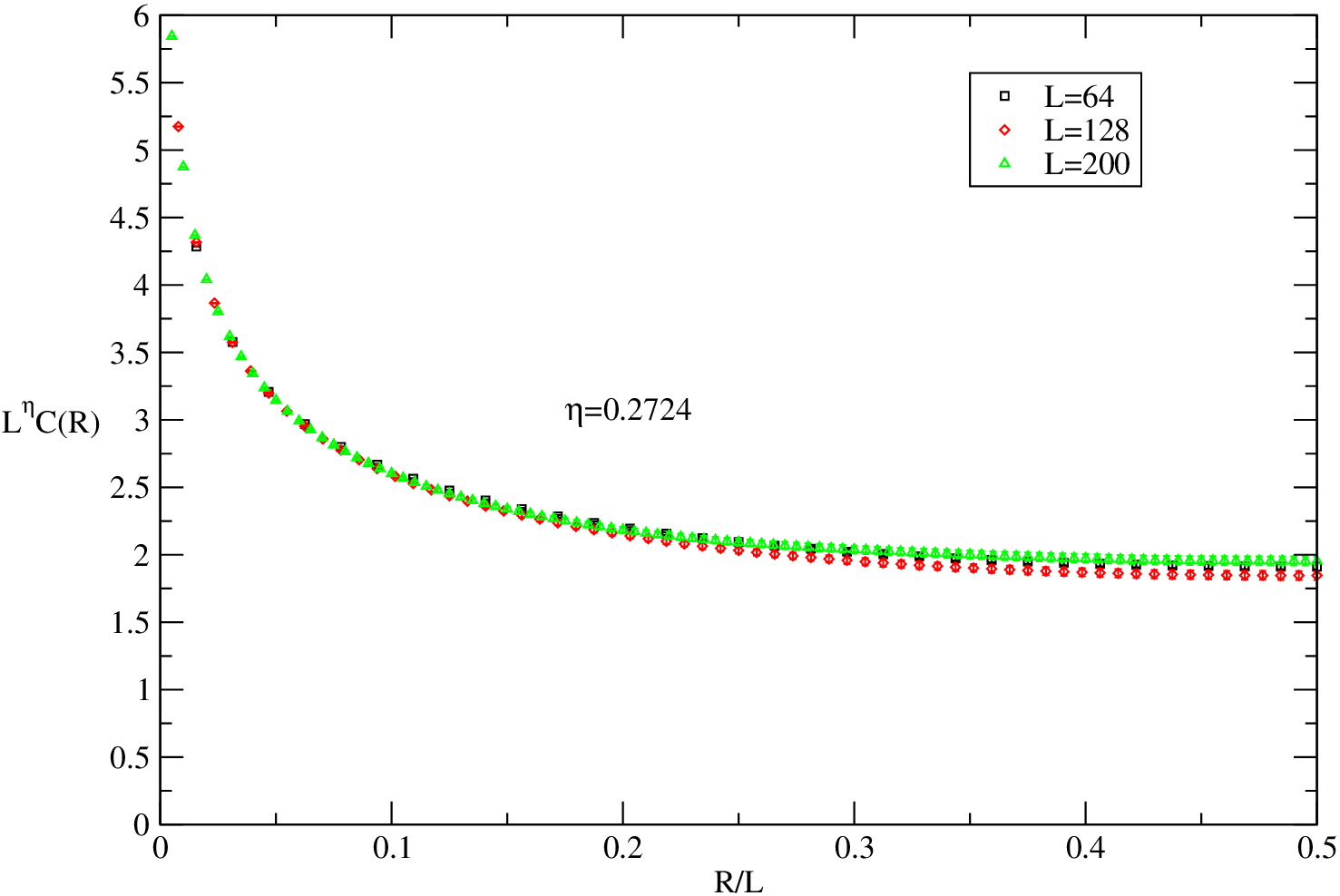}
\caption{(Top) $\eta_{\mbox{\scriptsize eff}}$ for $N_t=4$ on lattices 
with $L$=64, 128, 200 at $\beta=3.42$. For all lattices we fixed $R_0$=10.
Errors are determined by the jackknife method. (Bottom) $L^\eta C(R)$
versus $R/L$, with $\eta$ fixed at the central value of our determination
through the method of the effective $\eta_{\mbox{\scriptsize eff}}$
(see the text).}

\label{eta_4}
\end{figure}

\begin{figure}[tb]
\centering
\includegraphics[width=13cm]{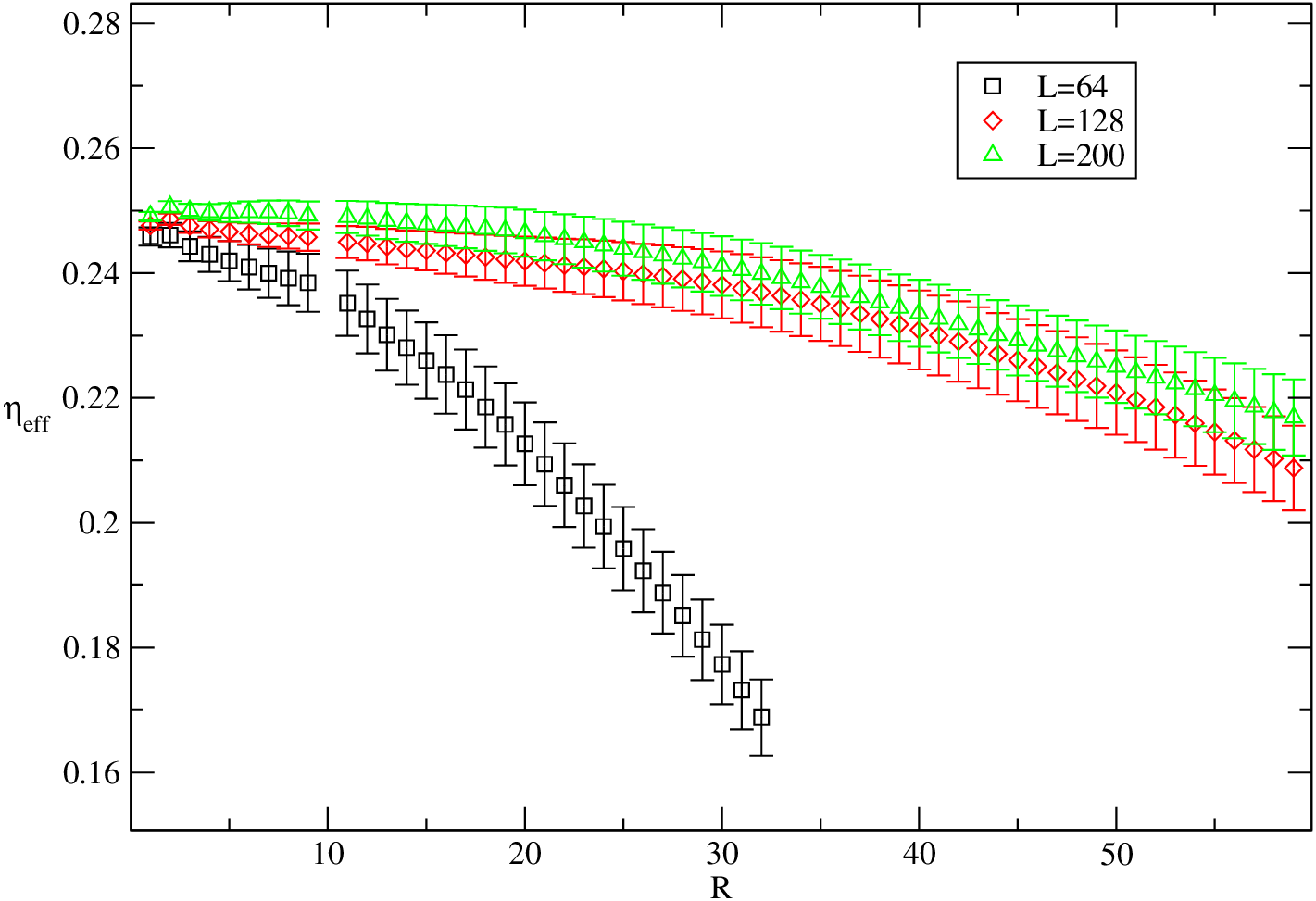}

\vspace{1.5cm}
\includegraphics[width=13cm]{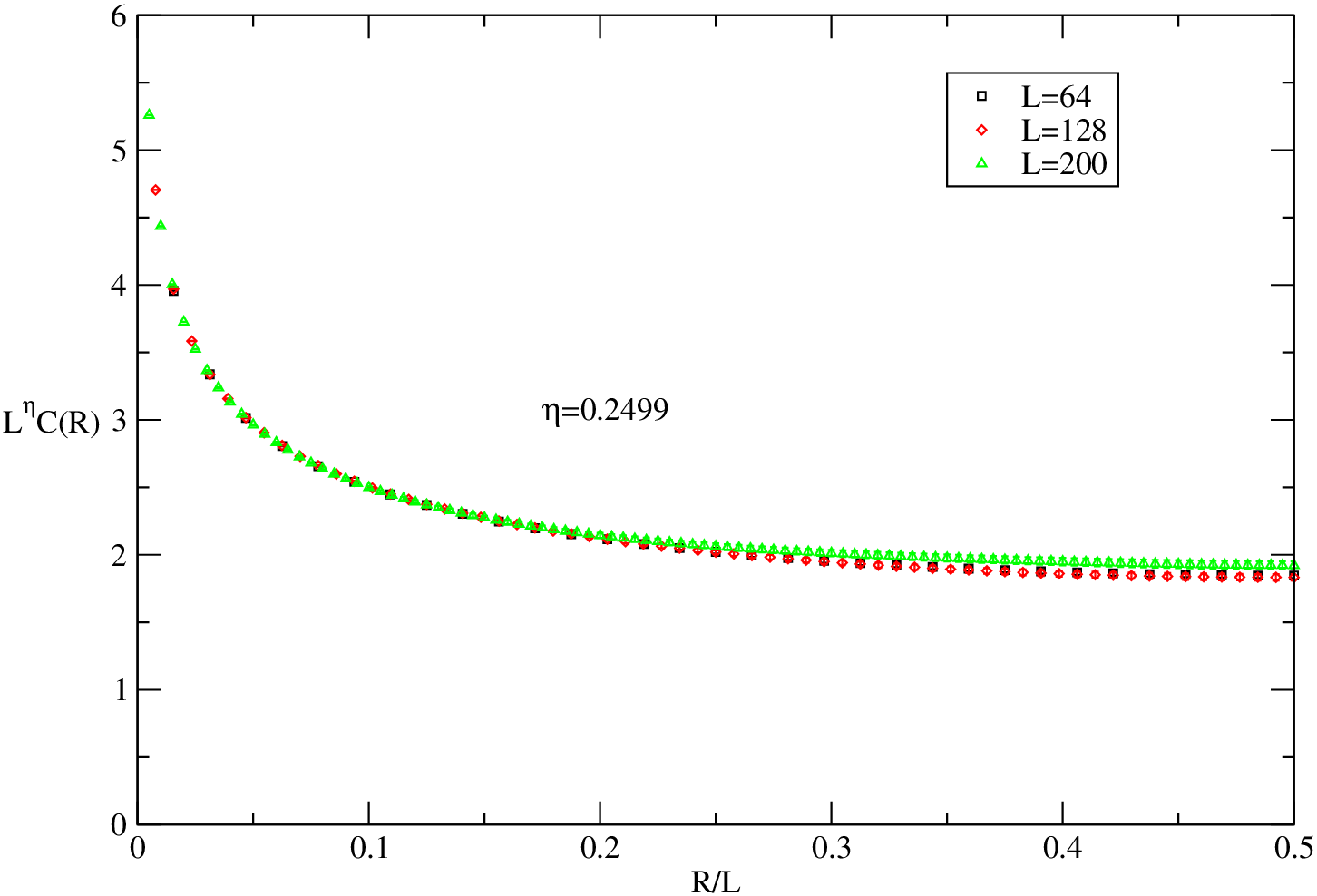}
\caption{(Top) $\eta_{\mbox{\scriptsize eff}}$ for $N_t=8$ on lattices 
with $L$=64, 128, 200 at $\beta=6.38$. For all lattices we fixed $R_0$=10.
Errors are determined by the jackknife method. (Bottom) $L^\eta C(R)$
versus $R/L$, with $\eta$ fixed at the central value of our determination
through the method of the effective $\eta_{\mbox{\scriptsize eff}}$
(see the text).}

\label{eta_8}
\end{figure}

A more precise determination of the magnetic index can be achieved 
through the study of the point-point correlation function. In 
Figs.~\ref{eta_4}(top) and \ref{eta_8}(top) we show 
$\eta_{\mbox{\scriptsize eff}}(R)$ for three values of the spatial size $L$ 
for the cases of $N_t = 4$ and $N_t=8$, respectively.
Our estimated plateau values, taken from data at $L=200$, are 
\begin{eqnarray*}
\eta(\beta=3.42)&=&\eta_{\mbox{\scriptsize eff}}(N_t=4,R=2)=0.2724(11)\;, \\
\eta(\beta=6.38)&=&\eta_{\mbox{\scriptsize eff}}(N_t=8,R=3)=0.2499(11)\;.
\end{eqnarray*}

For $N_t=4$, $\eta$ overshoots by little the $XY$ universal value, while for 
$N_t=8$ it is in nice accord with it. The deviation for $N_t=4$ is most likely 
washed out by a fine tuning of the critical coupling within its error bars.

One can observe, moreover, that the shape of the curve of values of
$\eta_{\mbox{\scriptsize eff}}(R)$ changes qualitatively in the same
way when the thermodynamic limit is approached for $N_t=1$ and $N_t=8$, 
while it has a different behavior for $N_t=4$. This may be an indication that 
for $N_t=1$ and $N_t=8$ at the $\beta$'s chosen for the simulation
the system is in the same phase ($\beta > \beta_c$), {\it i.e.} 
correlators have the same behavior.
 
Figs.~\ref{eta_4}(bottom) and~~\ref{eta_8}(bottom) show the correlation 
function $C(R)$ rescaled by $L^{-\eta}$ in units of $R/L$, with $\eta$
fixed at the central value of our determinations ($\eta=0.2724$ for $N_t=4$ 
and $\eta=0.2499$ for $N_t=8$); one can see that data from different lattices 
fall on top of each other over a wide range of distances. 

In summary, essential scaling is verified both for $N_t = 4$ and 8, thus
indicating that indeed the occurring transitions are compatible with BKT.
Moreover data point to values of the thermal and magnetic critical indices 
of the 2$d$ $XY$ universality class. This leads us to conclude that for $N_t=4$
and 8 the 3$d$ U(1) LGT at $\beta_s$=0 belongs to the 2$d$ $XY$ 
universality class and this supports the conjecture that the same holds, in 
general, for any $N_t$ at $\beta_s=0.$ 

Since we do not study the correlation length, we are not allowed to rule out
the possibility that it keeps finite and the transition is therefore first 
order. To this aim, we have performed a fit to the pseudo-critical couplings 
with the first order law
\be
\beta_{pc}(L)=\beta_c+\frac{B}{L^2}\;,
\label{first}
\ee
finding
\begin{eqnarray}
\beta_c(N_t=4)&=&3.245(3)\;, \;\;\;\;\; B(N_t=4)=-500(30)\;, \;\;\;\;\;
(\chi^2/\mbox{d.o.f.}=2.1) \nonumber 	\\
\beta_c(N_t=8)&=&5.852(8)\;, \;\;\;\;\; B(N_t=8)=-1300(100)\;, \;\;\;\;\;
(\chi^2/\mbox{d.o.f.}=0.6) \quad . 
\label{bc_first_4_8}
\end{eqnarray}
Looking at the $\chi^2$/d.o.f., one can argue that for $N_t=4$ first order 
should be ruled out, whereas $N_t=8$ is compatible with first order 
scaling.~\footnote{The same conclusion can be reached by studying the 
scaling with the lattice size of the peak of the Polyakov loop susceptibility.}
This can be due to the limited volumes ($L\leq 150$) considered for $N_t=8$
and to the larger error bars in the determinations of the $\beta_{pc}$'s with
respect to the $N_t=4$ case. However, for $N_t=8$ the good agreement
between the numerical result for the magnetic critical index and the
corresponding value in the $2d$ $XY$ model supports the claim that, even
for this $N_t$, the transition is BKT.

\section{Conclusions and outlook}

The purpose of this paper has been to study the critical behavior of 3$d$ U(1) 
LGT at finite temperatures, through the formulation on an asymmetric lattice.
While the theory at zero-temperature is always in the confined phase, at finite
temperatures it undergoes a deconfinement phase transition, just as it happens 
for 4$d$ QCD. Analytical results from the high-temperature expansion suggest 
that this transition is of BKT type, but compelling numerical evidence is 
missing that indeed critical indices of 3$d$ U(1) LGT coincide with those of 
the 2$d$ $XY$ model. 

This paper is the first step in the construction of the phase diagram of 3$d$ 
U(1) LGT in the $(\beta_t,\beta_s)$-plane, where $\beta_s$ ($\beta_t)$ is the 
spatial (temporal) coupling. In particular, we restricted ourselves to the case
$\beta_s=0$ and, by means of numerical Monte Carlo simulations on a 
dimensionally reduced effective theory, found evidence that the theory belongs 
indeed to the same universality class of the 2$d$ $XY$ model. The key 
observations have been the appearance of essential scaling and the agreement 
of the magnetic critical index $\eta$ with that from the 2$d$ $XY$ model.

The next step is the extension of the numerical procedure established in this 
paper to the general case of $\beta_s \neq 0$.

\vspace{1.0cm} \noindent
{\Large \bf Acknowledgment} \vspace{0.5cm}

O.B. thanks for warm hospitality the Dipartimento di Fisica dell'Universit\`a 
della Calabria and the INFN Gruppo Collegato di Cosenza, where the idea of 
this investigation came up. A.P. is grateful to the Department of Nuclear 
Physics of the Technical University of Vienna for hosting him during the final 
stages of the work.

\end{document}